\documentclass{aa}  

\usepackage{graphicx,lscape}
\usepackage[varg]{txfonts}
\usepackage{natbib}
\usepackage[english]{babel}

\usepackage{natbib,twoopt}

\newcommand{\ltapprox}{\raisebox{-0.5ex}{$\,\stackrel{<}{\scriptstyle\sim}\,$}}

\usepackage{color}
\usepackage{amstext}

\begin{document}

\title{MUSE observations of the lensing cluster Abell 1689}



 \author{D.~Bina\inst{1,2}
          \and
R.~Pell\'o\inst{1,2} \and
J.~Richard\inst{3}  \and
J.~Lewis\inst{1,2} \and
V.~Patr\'icio\inst{3}  \and
S.~Cantalupo\inst{4} \and
E.~C.~Herenz\inst{5} \and
K.~Soto\inst{4} \and
P.~M.~Weilbacher\inst{5} \and
R.~Bacon\inst{3}  \and
J.~D.~R.~Vernet\inst{6} \and
L.~Wisotzki\inst{5} \and
B.~Cl\'ement\inst{3}  \and
J.~G.~Cuby\inst{7} \and
D.~J.~Lagattuta\inst{3}  \and
G.~Soucail\inst{1,2} \and
A.~Verhamme\inst{3,8} 
          }

\institute{
Universit\'e de Toulouse; UPS-OMP; IRAP; Toulouse, France\\
\and
CNRS; IRAP; 14, avenue Edouard Belin, F-31400 Toulouse, France\\
\email{dbina@irap.omp.eu,rpello@irap.omp.eu}
\and
Univ Lyon, Univ Lyon1, Ens de Lyon, CNRS, Centre de Recherche Astrophysique de Lyon UMR5574, 
F-69230, Saint-Genis-Laval, France \\
\email{johan.richard@univ-lyon1.fr}
\and
ETH Zurich, Institute of Astronomy, Wolfgang-Pauli-Str. 27, CH-8093 Zurich,
Switzerland \\
\email{cantal@ucolick.org}
\and
Leibniz-Institut für Astrophysik Potsdam (AIP), An der Sternwarte 16, D-14482
Potsdam, Germany \\
\email{cherenz@aip.de}
\and
European Southern Observatory, Karl Schwarzschild Straße 2, 85748, Garching,
Germany \\
\email{jvernet@eso.org}
\and
Aix-Marseille Universit\'e, CNRS, LAM (Laboratoire d'Astrophysique de
Marseille) UMR 7326, 13388, Marseille, France \\
\email{jean-gabriel.cuby@lam.fr}
\and
Geneva Observatory, University of Geneva, 51 Chemin des Maillettes, 1290
Versoix, Switzerland  \\
\email{Anne.Verhamme@unige.ch}
}

\date{Received ; accepted }

\abstract
{This paper presents the results obtained with the Multi Unit Spectroscopic
Explorer (MUSE) for the core of the lensing cluster Abell 1689, as part of MUSE's 
commissioning at the ESO Very Large Telescope.}
{Integral-field observations with MUSE provide a unique view of the central 1 
$ \times$ 1 arcmin$^{2}$ region at intermediate spectral resolution in the
visible domain, allowing us to conduct a complete census of both 
cluster galaxies and lensed background sources.}
{We performed a spectroscopic analysis of all sources found in the MUSE data
cube. Two hundred and eighty-two objects were systematically extracted from the cube based on a 
guided-and-manual approach. We also tested three different tools for the automated
detection and extraction of line emitters. 
Cluster galaxies and lensed sources were identified based on their
spectral features. We investigated the multiple-image configuration for all
known sources in the field.
}
{Previous to our survey, 28 different lensed galaxies displaying 46 multiple
images were known in the MUSE field of view, most of them were
detected through photometric redshifts and lensing
considerations. Of these, we spectroscopically confirm 12 images based on
their emission lines, corresponding to 7 different lensed galaxies between $z =$ 0.95 and 5.0.
In addition, 14 new galaxies have been spectroscopically identified in this area thanks to
MUSE data, with redshifts ranging between 0.8 and 6.2.  
All background sources detected within the MUSE field of view correspond to 
multiple-imaged systems lensed by A1689. 
Seventeen sources in total are found at $z \ge 3$ based on their Lyman-$\alpha$
emission, with Lyman-$\alpha$ luminosities ranging between 
40.5$\lessapprox$log(Ly$\alpha$)$\lessapprox$42.5 after correction for
magnification. This sample is particularly
sensitive to the slope of the luminosity function toward the faintest end.
The density of sources obtained in this survey is consistent with
a steep value of $\alpha \le -1.5$, although this result still needs
further investigation.
}
{These results illustrate the efficiency of MUSE in characterizing
lensing clusters on one hand and in studying faint and distant populations
of galaxies on the other hand. In particular, our current survey of lensing
clusters is expected to provide a unique census of sources responsible for the
reionization in a representative volume at z$\sim$4-7.
} 


   \maketitle
%

\section{Introduction}
\label{intro}

   Clusters of galaxies are recognized to be the most efficient gravitational lenses in the
Universe
\citep[see, e.g.,][and the references therein]{2011A&ARv..19...47K}. 
For almost 30 years since their discovery, they have allowed the
astrophysical community to detect extremely faint and distant sources that
would otherwise be inaccessible. Strong magnification in the core of
lensing clusters has opened a brand-new type of observational approach
to characterize the most distant or faint sources while reducing the selection bias
in luminosity. Many high-redshift sources have been
identified and studied using this approach during the last years 
\citep[see, e.g.,][]{2012Natur.489..406Z, 
2013ApJ...762...32C, 
2015A&A...575A..92L, 
2015ApJ...800...18A}. 
Cluster lenses have also contributed to constrain the luminosity function for
galaxies at high-z down to the faintest limits, often through dedicated surveys  
\citep{2014ApJ...792...76B, 
2015ApJ...799...12I, 
2015ApJ...800...18A}. 
Using lensing clusters as efficient gravitational telescopes requires a
detailed knowledge of the mass distribution, which in turn depends on the
correct identification of multiply imaged systems in the cluster core 
\citep{2011MNRAS.410.1939Z, 
2014MNRAS.444..268R}. 

   This paper presents the results obtained with the Multi Unit Spectroscopic Explorer 
\citep[MUSE][]{2010SPIE.7735E..08B} 
of the core of the lensing cluster Abell 1689 (hereafter A1689), 
as part of MUSE's\ commissioning at the ESO Very Large Telescope. 
This cluster has been extensively
studied during the past 20 years because it is one of the most efficient
strong-lensing clusters currently known. Previous analysis of A1689 
discovered a large number of multiple images.  
\citet[][hereafter C10]{2010ApJ...723.1678C} 
used 135 images of 42 galaxies identified within the central $\sim$400 kpc
diameter region to constrain their mass model. 
One of the most spectacular cases of a strong-lensing configuration are the Sextet Arcs,
discovered in this cluster by
\citet{2007ApJ...665..921F}, 
and produced by a $z = 3.038$ Lyman-break galaxy (LBG). 
The large number of multiple images identified in this field has allowed a
reliable mass reconstruction for this cluster, making it one of the best-known
gravitational telescopes 
\citep[see, e.g.,][]{2007ApJ...668..643L, 
2008MNRAS.386.1169T, 
2010ApJ...723.1678C}. 
For this reason, A1689 has become a privileged line of sight along which to
study the properties of faint and distant samples of background galaxies 
\citep[][]{2014ApJ...780..143A}. 

   Integral-field observations of A1689 with MUSE provide a unique view of the
central 1 $\times$ 1 arcmin$^{2}$ region at intermediate spectral resolution in the
visible domain, allowing us to conduct a complete census of both 
cluster galaxies and lensed background sources, in particular in the $z \sim
2.7 - 6.5$ domain. Our performance assessment of MUSE in the field of A1689
confirms the idea that this instrument is particularly well suited for the
study of faint and distant populations of galaxies
\citep[see also][]{2015MNRAS.446L..16R, 
2015A&A...575A..75B}.  

   The outline of this article is as follows. 
In Sect.\ \ref{observations} we describe the observations carried out on
the field of A1689 and provide a summary of the data
reduction and analysis process. Section\ \ref{cluster} addresses the properties				
of the cluster galaxy populations as seen by MUSE. The identification of
background sources lensed by A1689 is discussed in Sect.\ \ref{images}, in				
particular the spectroscopic confirmation of known multiply imaged sources as
well as the discovery of new multiple systems. Implications of these
results on the mass modeling of A1689 are also discussed in this section. 
Conclusions and perspectives are given in
Sect.\ \ref{Conclusions}. Throughout this paper, we adopt a 						
$\Lambda$-CDM cosmology with $\Omega_{\Lambda}$ = 0.7, $\Omega_{m}$ = 0.3
and $H_{0}=$ 70\ km\ s$^{-1}$\ Mpc$^{-1}$. Magnitudes are given in the AB
system \citep{1983ApJ...266..713O}. 
All quoted redshift measurements are based on vacuum rest-frame
wavelengths unless otherwise specified.

\section{From MUSE data to source extraction}
\label{observations}

\subsection{Observations and data reduction}

   The core of A1689 
\citep[$z =$ 0.1847,][]{1990ApJS...72..715T} 
was targeted with MUSE during its first commissioning run on February 9, 
2014 (ESO program 60.A-9100(A)). We observed a single field of view,
centered on $\alpha$=13:11:30.5, $\delta$=-01:20:41.5 (J2000) 
and oriented at PA=0\degr,  
covering a field of view of about 1 $\times$ 1
arcmin$^{2}$ ($\sim$185$\times$185 kpc at the redshift of the cluster).
The total exposure time was $\sim$1.83 hours, divided into six
individual exposures of 1100 seconds of integration time 
with a small linear dither pattern of 0.2\arcsec and no rotation between each
exposure. Observations were obtained in the nominal WFM-NOAO-N mode of MUSE, 
in good seeing conditions ranging between $\sim$0.51\arcsec and 0.87\arcsec, 
as measured by the DIMM seeing monitor, and 
between $\sim$0.9\arcsec and 1.1\arcsec (FWHM) as measured from MUSE 
data cubes in the reconstructed white-light images.  

   MUSE data were reduced using the MUSE Data Reduction System V1.0
\citep{2012SPIE.8451E..0BW, 
2014ASPC..485..451W}, 
including bias, dark, flat-fielding, and geometrical corrections, basic sky
subtraction, wavelength and flux calibrations, and astrometry. 
The six individual exposures were reduced and combined into a final datacube. 
Sky subtraction and correction for sky residuals were
performed with the ZAP software 
\citep[Zurich Atmoshere Purge][]{2016ascl.soft02003S}. 
Several versions of the MUSE datacube were performed, with different
choices for the combination parameters and the correction for sky residuals. 
Flux calibration and telluric correction was
achieved based on observations of the standard star HD49798 (O6 star). The
final datacubes extend between 4750 and 9351$\AA$, that is, 3681 spectral pixels
with a scaling of 1.25$\AA$/pixel and a spatial scale of
0.2\arcsec/pixel. The spectral resolving power varies between $\sim$2000 in the
blue and 4000 in the red. The seeing measured on the combined exposure is
$\sim$0.6\arcsec FWHM at 7300$\AA$. 

   In addition to MUSE data, we used archive observations obtained by the Hubble
Space Telescope (HST) with ACS in the optical bands (F475W, F625W, F775W,
F850LP; PID 9289, PI: H. Ford). These deep images reach a limiting magnitude
of AB$\sim$28.2 in the F775W filter (5$\sigma$ in a 0.2\arcsec radius
aperture)
\citep[see also][]{2014ApJ...780..143A}. 
Sources were extracted using two different approaches: a guided extraction
based on the detection on HST images, and a blind detection of line-emitters
using different automated tools. 
A color image of the HST field of view showing the footprint of the MUSE field
is displayed in Fig.\ \ref{finding_chart},
together with the lensed sources that are spectroscopically identified
or confirmed in this work.
%

\subsection{Extraction of HST-detected sources}
\begin{figure} [h]
\begin{center}
\includegraphics[width=0.48\textwidth]{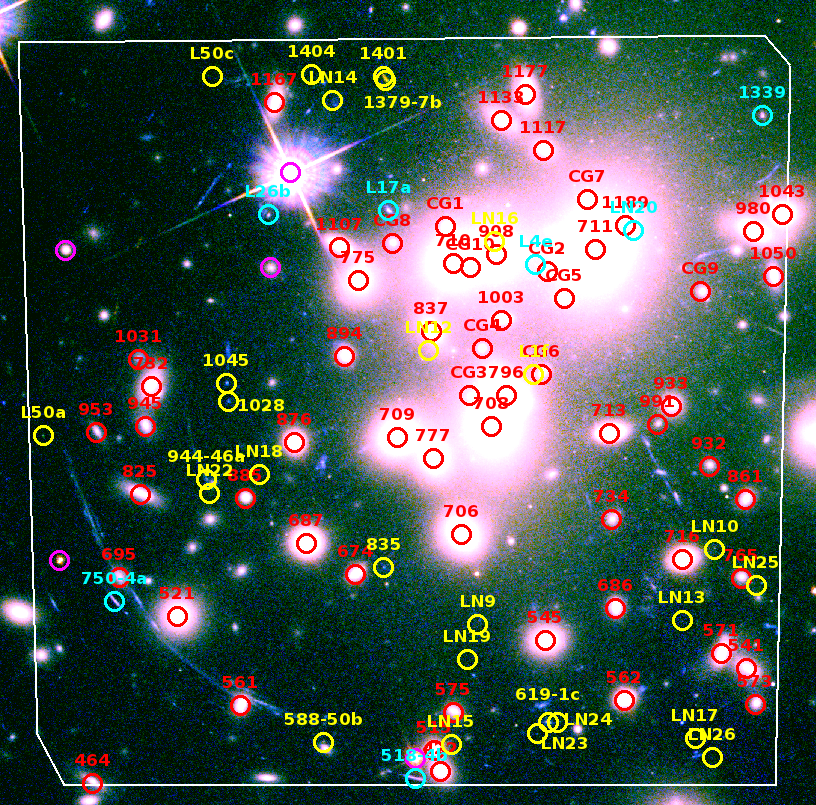}
\caption{Color image of A1689 obtained by combining F475W, F775W, and F850LP
HST/ACS images using a logarithmic scale. The footprint of the MUSE field is displayed together with
the sources spectroscopically identified in this field: cluster galaxies
(red), stars (magenta), background galaxies at $z \le 3$ (cyan), and $z \ge 3$
(yellow). Cluster galaxies that were manually extracted are labeled CG. Objects labeled
LN correspond to either line emitters or multiple images lacking a counterpart in the
reference catalog based on the F775W image (see text for details). 
North is up, east is left. 
A tentative spectroscopic redshift was also obtained for the long
southeast arcs in this field, namely 8a and 8b in C10 (see discussion in 
Sect.\ \ref{old_sources}).
At the redshift of the
cluster, the MUSE field is equivalent to $\sim$185$\times$185 kpc.
} 
\label{finding_chart}
\end{center}
\end{figure}

Sources were extracted using two different approaches. First, a catalog of
objects was built with SExtractor package version 2.8 \citep{Sex} based on the 
detection on the F775W image. Two hundred forty-five objects were detected in this way within the
MUSE field of view, with Kron-like magnitudes (SExtractor MAG\_AUTO) 
ranging between 17.0 and 28.1. 
Before extracting background sources in the central region from
the MUSE cube, background residuals and continuum
emission from bright cluster galaxies (hereafter BCGs) 
were fitted and removed from the MUSE cube 
with median filtering in both wavelength and spatial directions.
The BCG area hereafter refers to the central region, which is
strongly contaminated by
the bright galaxies and displayed in a pink color-scale in Fig.\ \ref{finding_chart}.
When building the catalog, ten bright sources in the cluster
core that were missed by SExtractor because of severe crowding, were manually added. 
In addition, 23 multiple images from
\citet{2010ApJ...723.1678C} were missing from the SExtractor catalog for
various reasons, mainly because they are located within or close to the central
halo. They were also added to the catalog, as well as the four line emitters that were
serendipitously discovered after manual inspection of the cube. This process
provided 282 objects within the MUSE field of view. 1D spectra for all these
sources were systematically extracted from the MUSE cube using a fixed
aperture corresponding to 1.5\arcsec diameter. 

   Based on this guided-and-manual approach, a secure redshift could be determined
for all sources up to MAG\_AUTO m(775W) $\sim$ 21.5, most of them being cluster
galaxies or stars identified based on their continuum emission. 
Redshifts were first measured based on manual inspection, followed by template
fitting. In total, the spectroscopic sample obtained in this way over the entire
photometric catalog yields 63 cluster galaxies (49 of them with m(775W) $<$ 21.5), 
5 stars, and 24 galaxies behind the 
cluster, 17 of them corresponding to sources at $z \ge 3$ detected by their
Lyman-$\alpha$ emission. The majority of these $z \ge 3$ identifications 
were verified manually and by lensing (see below). 

\subsection{Blind extraction of line emitters}
The second approach was the automated (blind) detection and extraction of
line emitters and continuum based on three main tools that are developed 
within the MUSE consortium, namely CubExtractor, LSDCat, and MUSELET. 
CubExtractor (Cantalupo in prep.)
detects and extracts sources as connected voxels (pixels in the cube) above a given 
pixel-to-pixel signal-to-noise ratio (S/N) threshold, after 3D Gaussian filtering of the MUSE datacube.
LSDCat (Herenz in prep.) is optimized for the extraction of compact narrow
emission-line objects using a matched-filtering approach. 
MUSELET (for MUSE line-emission tracker, Richard in prep.) is a simple 
SExtractor-based tool that extracts sources displaying
line-emission, regardless of whether they are linked with continuum emission.
All these tools have been optimized for the extraction of line
emitters in
different environments. Using them in this crowded field was particularly 
challenging. 

A systematic manual inspection was needed to exclude spurious
detections, most of them coming from sky-line residuals or
imperfect correction for cosmic rays (sometimes showing extended structure in
the final cube). During this phase, we used three different versions of the
MUSE datacube, with different choices in the parameters used for background
subtraction, sky-line residuals and cosmic-ray corrections. Reliable
emission lines are expected to be detected in all the cubes. The percentage of spurious 
detections depends on the flux level of the lines and the parameters used for
the automated detection, typically ranging from a few percent for the
brightest sources to as high as $\sim$75\% for the faintest ones 
(e.g., flux $\sim$5$\times$10$^{-18}$ erg/s/cm$^{2}$). 

   These tools used in various combinations allowed us to detect
14 new emission-line objects, most of them lacking a clear counterpart in the
HST/ACS images. All line emitters extracted based on the guided-and-manual
approach and discussed in the previous section were
detected by one of the automated tools. All sources retained in 
Sect.\ \ref{images} were found in all different versions of the MUSE data
cubes (with and without background residuals and continuum emission
subtracted). All of them yield multiple-image configurations 
in good agreement with lensing expectations. 
Based on our experience, the automated procedure was especially useful to find faint emission-line
objects, with a typical flux of a few 10$^{-18}$ erg/s/cm$^{2}$, in particular
those lacking a clear detection on HST/ACS images. 

\section{Cluster galaxies}
\label{cluster}

\begin{figure}
\begin{center}
\includegraphics[width=0.48\textwidth]{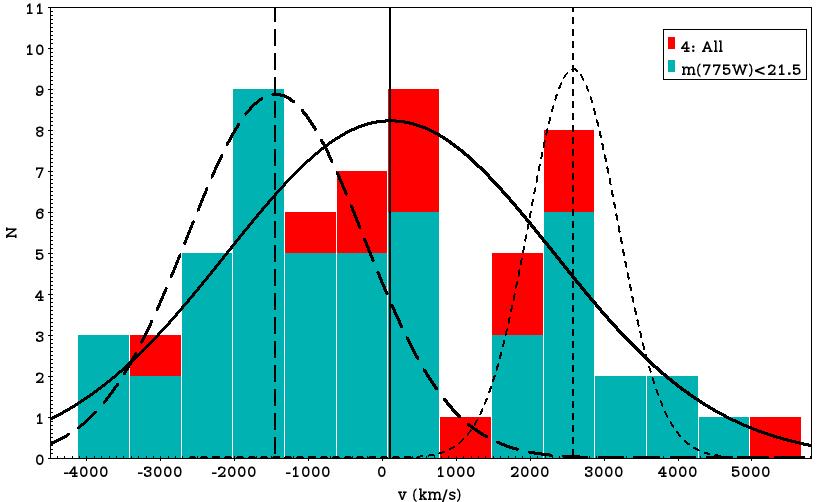}
\caption{Histogram showing the velocity distribution for the 63 cluster
galaxies found in the MUSE field of view with respect to the centroid for the
whole distribution ($z =$ 0.1891). 
Solid lines display the best Gaussian fit to the whole sample. 
Long and short dashed lines display the best fit to the main cluster component (low-z) 
and the northeast group (high-z), respectively.
}
\label{histo_BCG_vel}
\end{center}
\end{figure}

   The spectroscopic sample obtained in the MUSE field includes 63 cluster
galaxies with secure redshifts. All of them exhibit early-type spectra,
excepted for two emission-line galaxies. Forty-nine of them are bright, that is, have 
m(775W) $<$ 21.5. They constitute the majority of the sample within the
spectroscopic completeness. Redshifts were measured based on the 
fit of several absorption lines with high S/N (Balmer series, MgI, Gband,
etc.). The mean and the median redshifts
for this sample are $z =$ 0.1891 and 0.1889,
respectively ($z =$ 0.1883 and 0.1875 when the sample is limited to the
completeness level at m(775W) $\le$ 21.5). When adopting the mean redshift for
the whole sample as the heliocentric reference, the velocity dispersion is found to be
$\sigma =$ 2214 $\pm$ 38 km/s,
in good agreement with previous findings in the
literature that used more galaxies and wider regions (typically a few
Mpc) \citep[see, e.g.,][]{1990ApJS...72..715T, 
1991ApJS...77..363S} 
This value also agrees well with the largest spectroscopic survey
conducted so far on this cluster by \citet {2004ogci.conf..183C}, 
which included 525 cluster members and displayed a velocity dispersion profile
decreasing from $\sim$2100 km/s in the center to 1200 km/s at
$\sim$3$h^{-1}$Mpc from the core. Figure\ \ref{histo_BCG_vel} displays the
velocity distribution found in the MUSE field of view with respect to the
centroid of the whole distribution, and the best Gaussian fit. The
velocity dispersion does not depend on the sample and the reference used
(i.e., when the center of mass is computed based 
on either the whole sample or the bright sample alone), all $\sigma$ values
are the same within a $\pm$0.5\% interval. 

    However, Fig.\ \ref{histo_BCG_vel} displays a
more complex distribution than the smooth Gaussian distribution found for
this cluster on larger scales. \citet{2004ogci.conf..183C}, 
found no evidence for any substructure at $\sim$1$h^{-1}$Mpc scales, 
but a group of bright galaxies lies at $\sim$350 kpc
northeast of the cluster center, therefore outside the MUSE field of view, but is still
visible in the velocity distribution. This group is 
typically interpreted as evidence for a non-relaxed state in the cluster
center that is possibly due to an ongoing merger \citep {2007ApJ...668..643L}. 
This close group skews the redshift distribution toward slightly higher redshifts
($z \ge 0.195$) and broadens the redshift distribution in the cluster core. 
Figure\ \ref{histo_BCG_vel} also displays the best fit found when considering two
independent components, the main cluster component with 
$\sigma$ = 1200 $\pm$ 51 km/s,
and the group component with 
$\sigma$ = 600 $\pm$ 68 km/s.
The velocity offset between the two subclumps is 4030 km/s.
As a consequence of this effect, the mean redshift is slightly biased toward  
a higher value than in previous findings, that is, a value close to $z =$ 0.188 compared to $z =$ 0.184 in the references given above. The redshift of the
main cluster component is fully consistent with $z =$ 0.184. 
There is no obvious trend in the spatial distribution of galaxies belonging to
the group across the MUSE field of view. 

We have also checked the consistency between our redshifts for 
galaxies in the center of the cluster with respect to existing data,
being aware that spectroscopic targets were more sparsely sampled in previous surveys. 
Three galaxies of our sample are in common with \citet[][]{1990ApJS...72..715T} 
(708, 710, and 711 in Fig.\ \ref{finding_chart}), and four galaxies with 
\citet[][and private communication]{2004ogci.conf..183C}, 
(545, 713, 777, and 732 in Fig.\ \ref{finding_chart}). 
The average difference is consistent with zero: $<dz>=$0.0001$\pm$0.001. 

   The results above show that MUSE is able to catch the small-scale
substructure in particularly dense environments, a difficult task for standard
multi-slit specrographs. This result supports the method adopted by 
\citet[][hereafter L07]{2007ApJ...668..643L} 
to optimize the lensing mass-modeling in this cluster, 
based on observations by \citet{2004ogci.conf..183C} for the cluster dynamics 
and 32 multiply imaged systems known in this cluster at that epoch. They used 
a parametric reconstruction of the mass-distribution based on the
superposition of two dominant dark-matter clumps, with $\sigma$ = 1200 and 540 km/s,
respectively, to reproduce the large-scale component. A refined
version of the L07 model is used in this paper (see Sect.\ \ref{images}). 


\section{Lensed sources behind A1689}
\label{images}

   Previous to our survey, 28 different lensed galaxies displaying 46 multiple
images were known in the MUSE field of view, most of them based on photometric
redshifts together with lensing considerations 
\citep[see, e.g.,][]{2007ApJ...668..643L, 
2007ApJ...665..921F, 
2008MNRAS.386.1169T, 
2010ApJ...723.1678C}. 
Of these, we have been able to confirm 12 images corresponding to 7
different lensed galaxies between $z =$ 0.95 and 5.0. These systems are
described in Sect.~\ref{old_sources}. Fourteen new galaxies have
also been spectroscopically identified in this area thanks to 
MUSE data, with redshifts ranging between 0.8 and 6.2, as 
presented in Sect.~\ref{new_sources}.
The majority of the lensed galaxies has been identified based on their
Lyman-$\alpha$ emission. 

   The lensing model used to analyze these data is the
one presented in L07, 
based on Lenstool 
\citep[][]{2007NJPh....9..447J}, 
with some refinements to include the latest spectroscopic constraints
on multiple images available in the literature for this cluster 
(see also Richard et al. in prep.).
Table ~\ref{lens_model} provides the updated version of Table 3 in L07,
summarizing the main parameters used throughout this
paper for the lensing model. Dark matter halos are
modeled by a truncated pseudo-isothermal elliptical mass distribution scaled
to their masses, from large-scales (clumps 1 and 2, as discussed in 
Sect.~\ref{cluster}) to cluster-galaxy halos. The main contributions are listed in
Table ~\ref{lens_model} together with the basic scaling parameters used to
model the contribution of 192 additional early-type cluster galaxies. 
All images detected in this small field are expected to be
multiply imaged systems lensed by A1689. Our results below are fully consistent with
this expectation, as discussed in Sect.~\ref{model}. 

{\bf
\begin{table*}
\begin{center}
\caption{\label{lens_model} Summary of the main parameters defining the
lensing model. This table is an updated version of Table 3
in L07.}
\begin{tabular}{lccccccc}
\hline \hline
Mass Clump & RA (J2000) & DEC (J2000) & {\it e} & $\theta$ & r$_{core}$ &
r$_{cut}$ &  $\sigma_{0}$ \\ 
 & (1) & (1) & (2) & (3) & (kpc) &
(kpc) & (km.s$^{-1}$) \\
\hline
Clump 1 & 0.6 & -8.9 & 0.22 & 91.8 & 100.5 & 1515.7 & 1437.3 \\ 
Clump 2 & -70.0 & 47.8 & 0.80 & 80.5 & 70.0 & 500.9 & 643.2 \\
BCG      & -1.3 & 0.1 & 0.50 & 61.6 & 6.3 & 132.2 & 451.6  \\
Galaxy 1 & 49.1 & 31.5 & 0.60 & 119.3 & 26.6 & 179.6 & 272.8 \\
Galaxy 2 & 45.1 & 32.1 & 0.79 & 42.6 & 18.1 & 184.8 & 432.7 \\
Galaxy 3 & -28.8 & 55.2 & 0.15 & 57.5 & 0.17 & 62.9 & 162.3 \\
L$^{*}$ elliptical galaxy &  &  &  & & 0.15 & 18.2 & 159.6 \\
\hline
\end{tabular}
\tablefoot{\\
(1) Coordinates are in \arcsec with respect to the center of the field, as set
by L07, namely $\alpha$=13:11:29.51, $\delta$=-01:20:27.6 (J2000) \\
(2) The definition of ellipticity {\it e} is the same as for L07 \\
(3) The definition of the position angle $\theta$ is the same as for L07
}
\end{center}
\end{table*}
}

\begin{figure*}
\begin{center}
\includegraphics[width=0.95\textwidth]{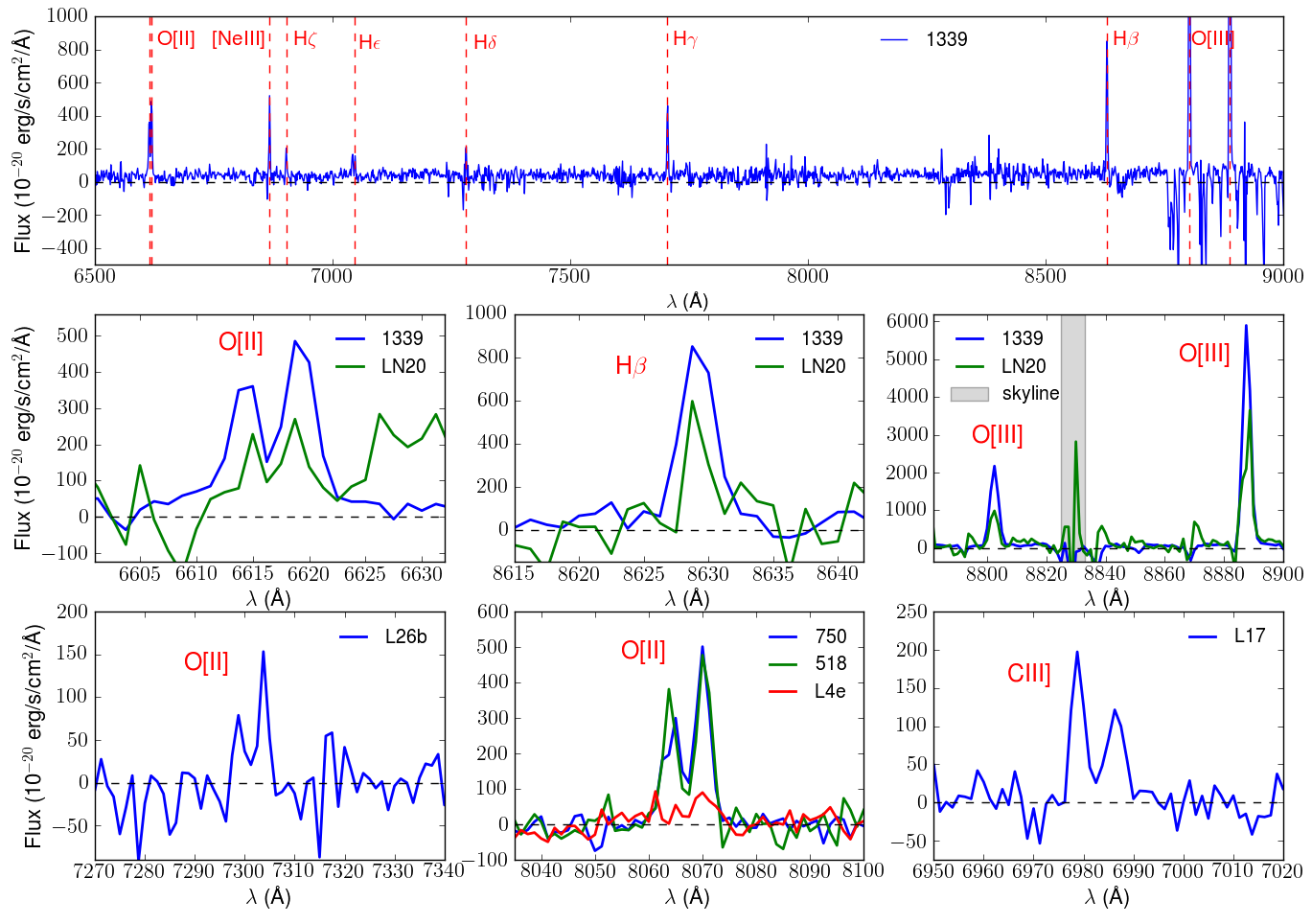}
\caption{Extracted spectra of all multiply imaged systems identified in the
MUSE field of view at $z \le 3$, ordered according to increasing redshift, following the
identifications in Table~\ref{catalog}. Continuum emission has been subtracted
in all cases. The flux for counter-image LN20, extracted within a BCG, has
been multiplied by a factor of 3 to improve readability. 
} 
\label{trombino_spectres1}
\end{center}
\end{figure*}

\begin{figure*}
\begin{center}
\includegraphics[width=0.80\textwidth]{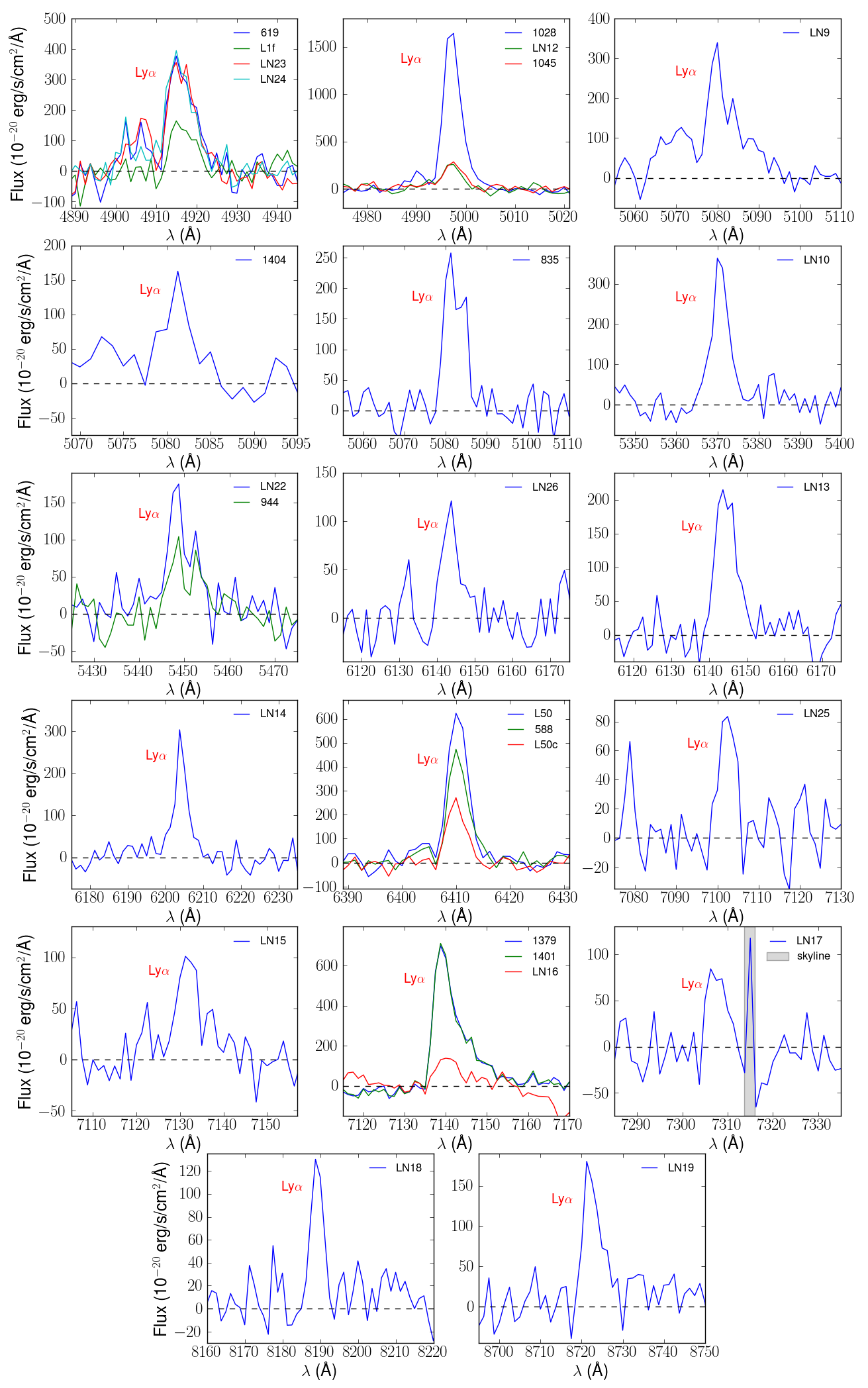}
\caption{Same as in Fig.\ \ref{trombino_spectres1} for multiply
imaged
systems at $z \ge 3$. The gray area corresponds to a strong emission-line residual.} 
\label{trombino_spectres2}
\end{center}
\end{figure*}

\begin{figure}
\begin{center}
\includegraphics[width=0.48\textwidth]{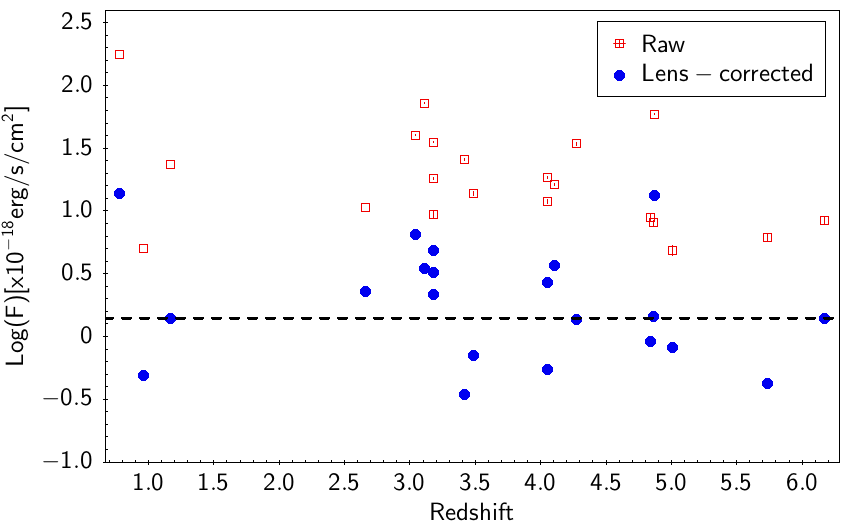}
\caption{Distribution of emission-line fluxes measured for the brightest
emission-lines versus redshift for the 21 sources found
behind A1689, with and without correction for lensing magnification. 
The lines used are given in Col. 3 of Table~\ref{catalog} (Lyman-$\alpha$ at
$z \ge 3$). The horizontal dashed line indicates the median lens-corrected flux
value. Multiply imaged sources are counted once in this diagram. 
} 
\label{plot_Flux_z}
\end{center}
\end{figure}

   Table ~\ref{catalog} presents the catalog of lensed images
or sources identified in this
field of view. Objects labeled LN in this table (Col. 1) correspond either to
line emitters or multiple images lacking a counterpart in the reference
catalog based on the F775W image. When the image or source was known before, its
previous identification is given in Col. 2.
%
Line fluxes were measured after subtracting the underlying continuum. 
Line profiles were fitted using MPDAF tools 
(Piqueras et al., in preparation).
A double-Gaussian profile was adopted to fit Lyman-$\alpha$ lines, combining two half Gaussian   
to better mimic the asymmetric profiles.
For Lyman-$\alpha$ emitters, we adopted the maximum of the line as reference for
redshift determinations.
Figures\ \ref{trombino_spectres1} and\ \ref{trombino_spectres2}
display the extracted spectra
for all images after subtracting underlying continuum, grouped by lensing systems. 
Figure\ \ref{plot_Flux_z} displays the distribution of
emission-line fluxes versus redshift measured for the brightest emission lines (usually
Lyman-$\alpha$) versus redshift for the 21 galaxies found
behind A1689. We note that multiply imaged sources are counted only once in this
diagram, and only the brightest line is shown for a given source. 
The median lens-corrected flux is $\sim$1.4$\times$10$^{-18}$ erg/s/cm$^{2}$. 
This illustrates the effective depth achieved in this survey
based on a relatively modest amount of exposure time. 


\subsection{Results on previously known systems}
\label{old_sources}

   Our MUSE survey has produced some interesting results on previously
known
systems in this field, in particular those exhibiting emission
lines. We summarize these below.  

\begin{itemize}

\item {\bf System 1 (619(1c) and L1f):}
One of the most spectacular sources in this field is the Sextet Arc,
corresponding to six separate images of a strongly lensed LBGs at $z =$ 3.038. 
\citep{2007ApJ...665..921F}. 
Two images of the source are present in the MUSE field of view, namely 1c and 1f. As shown by 
\citet{2007ApJ...665..921F}, 
this galaxy exhibits a complex behavior that is highlighted by magnification,
providing an enhanced spatial resolution. The spectrum is dominated by
absorption in image 1a, displaying a complex Lyman-$\alpha$ profile across the galaxy,
whereas a strong Lyman-$\alpha$ emission is found in image 1c. 
We here confirm the redshift found for 1c and provide an 
additional proof of the reality of this arc by confirming the Lyman-$\alpha$
emission of 1f. Before our survey, only a photometric redshift was available
for this image (see C10). In addition to this result, an extended emission is
also observed associated with 619(1c), with two main Lyman-$\alpha$ extensions labeled
LN23 and LN24 in Table~\ref{catalog}, as shown in Fig.\ \ref{System1}. Given
the location of 1f within the BCG region and its smaller magnification, the counter
images of LN23 and LN24 could not be detected. 

\begin{figure}
\begin{center}
\includegraphics[width=0.48\textwidth]{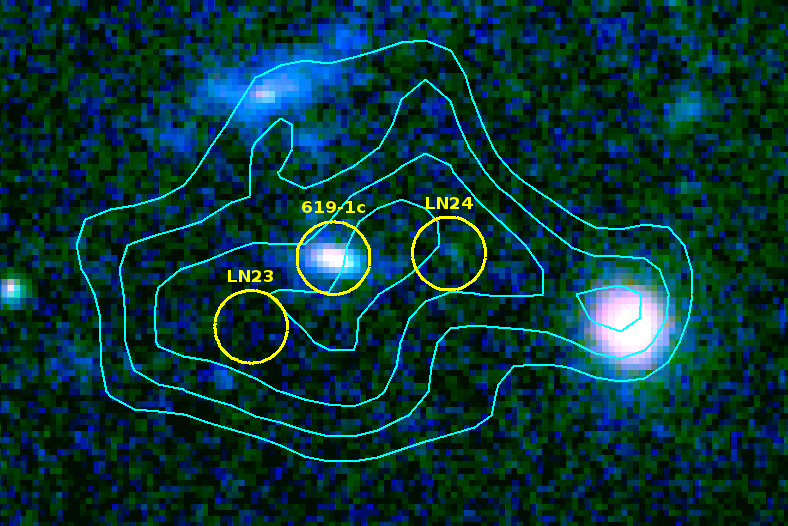}
\caption{Color image of A1689 obtained by combining F475W, F775W, and F850LP
HST/ACS images using a logarithmic scale, showing a
6.3\arcsec $\times$ 4.2\arcsec region around 619(1c)
and its extended emission. 
Circles have diameters of 0.60\arcsec, 
close to the FWHM seeing measured in the MUSE data.
Cyan isocontours display flux levels for
Lyman-$\alpha$ emission, from 0.5 to 2$\times$10$^{-19}$ erg/s/cm$^{2}$/pix$^{2}$, with 
bins of 0.5$\times$10$^{-19}$ erg/s/cm$^{2}$/pix$^{2}$. 
} 
\label{System1}
\end{center}
\end{figure}


\item {\bf System 4 (750(4a), 518(4b), and L4e):} 
This five-image system was described for the first time by L07,
who provided a first spectroscopic $z =$ 1.1, in agreement with the 
photometric redshifts derived by C10. A precise spectroscopic redshift was
used by \citet{2010Sci...329..924J}, 
fully consistent with the present value. 
We here spectroscopically confirm the system
as being the multiple image of a source at $z =$ 1.164, based on the detection of the same
identical emission-line doublet of [OII]3727
in the three images included in the MUSE field of view. 

\item {\bf System  7 (1379(7b), 1401, and LN16):} 
The multiply imaged system 7 was previously identified by 
\citet{2002ApJ...568..558F} 
and later discussed by
L07 and C10, 
who provided a photometric and lensing-motivated identification for 7b.
\citet{2002ApJ...568..558F} and 
\citet{2011MNRAS.413..643R} 
reported a redshift $z =$ 4.860 for the 7a component, which is
located outside the MUSE
field, based on the detection of Lyman-$\alpha$ emission for the former 
and the [OII]3727 in the near-IR for the latter. 
Here we confirm the nature of this system, with a spectroscopic redshift
$z =$ 4.874 for 7b, based on a strong emission-line identified
as Lyman-$\alpha$. The emission region seems to extend beyond the image seen on
HST data, toward the neighboring source 1401, 
as shown in Fig.\ \ref{System7}. 
We note that there is a difference in redshift
of $\sim$0.011 ($\delta$v$\sim$560 km/s)
between the 7b component reported here and the 
7a component (using air wavelength reference for all measurements to be consistent),
which cannot be easily explained. The extended
Lyman-$\alpha$ emission in this source might be sampled differently in the two
images. In addition, we confirm the
spectroscopic redshift of a counter-image located within the BCG area, labeled
LN16 in Table ~\ref{catalog}. Our results are fully consistent with a three-image
system, as suggested by previous authors. 

\begin{figure}
\begin{center}
\includegraphics[width=0.48\textwidth]{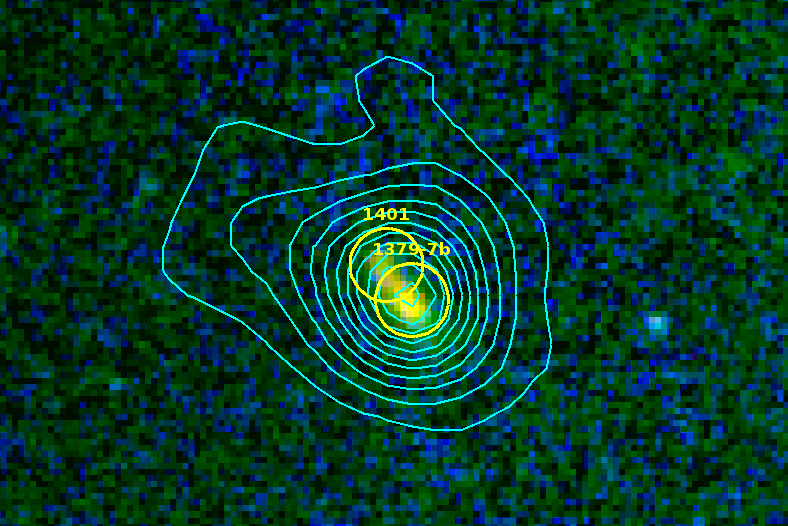}
\caption{Color image of A1689 obtained by combining F475W, F775W, and F850LP
HST/ACS images using a logarithmic scale, 
showing a 6.3\arcsec$\times$4.2\arcsec region around 1379 
and 1401. 
Circles have diameters of 0.60\arcsec, 
which is close to the FWHM seeing measured in the MUSE data.
Cyan isocontours display flux levels for
Lyman-$\alpha$ emission from 0.5 to 18$\times$10$^{-19}$ erg/s/cm$^{2}$/pix$^{2}$, with 
bins of 1 $\times$10$^{-19}$ erg/s/cm$^{2}$/pix$^{2}$.
} 
\label{System7}
\end{center}
\end{figure}

\item {\bf System 17 (L17a):} 
We spectroscopically confirm the redshift of the three-image system 17 in C10
as $z =$ 2.661, based on identifying the emission-line doublet of
CIII]1909. This value agrees well with the
photometric and lensing redshifts around $z =$ 2.7$\pm$0.5 estimated by 
L07 and C10. 

\item {\bf System 26 (L26b):} 
We spectroscopically confirm the redshift of the three-image system 26 in C10
as $z =$ 0.959, based on identifying the emission-line doublet of
[OII]3727. This value agrees excellently well with the photometric and
lensing redshifts $z =$ 0.9$\pm$0.2 estimated by C10. In addition, the counter-image 26c in C10,
which has a smaller magnification factor of $\sim$3.2, is also 
detected as a faint emission seen in the 3D MUSE cube, but it
is too faint to be 
successfully extracted. These results fully confirm the lensing system.  

\item {\bf System 46 (944(46a) and LN22):} 
This highly magnified double-image system was reported in C10 with uncertain
photometric redshifts for the two images. Here we measure a redshift $z =$ 3.483
based on an emission line identified as Lyman-$\alpha$. The two objects showing
this emission on the 1.5\arcsec diameter aperture are located side by side,
but the emission line is clearly centered on LN22, lacking a clear counterpart
in the HST/ACS images, as shown in Fig.\ \ref{System46}. 
Based on these measurements, it is more
likely a three-image system, with two images located within the MUSE field. 
The fainter counter-images are expected to be located within the central
(noisy) region, with fluxes ($\sim$1$\times$10$^{-18}$ erg/s/cm$^{2}$)
that are too
faint to be detected with a significant S/N in this area. 

\begin{figure}
\begin{center}
\includegraphics[width=0.48\textwidth]{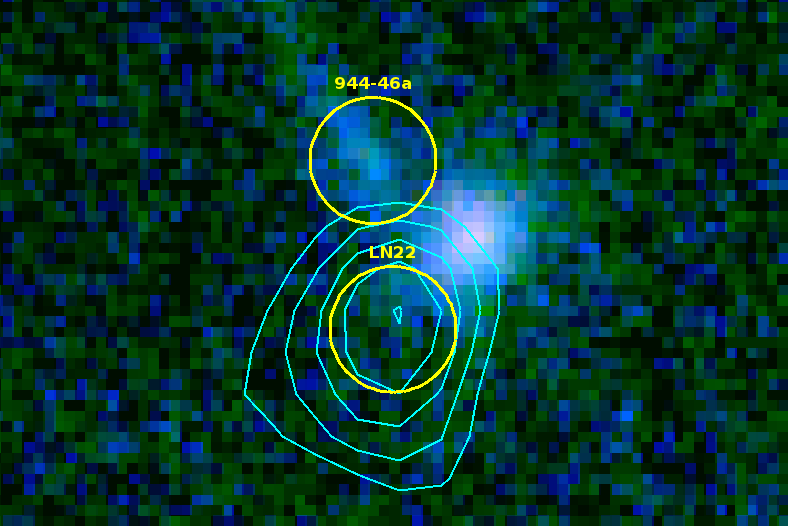}
\caption{Color image of A1689 obtained by combining F475W, F775W, and F850LP
HST/ACS images using a logarithmic scale, 
showing a 3.7\arcsec $\times$ 2.45\arcsec region around 944 
and LN22. 
Circles have diameters of 0.60\arcsec, 
which is close to the FWHM seeing measured in the MUSE data.
Cyan isocontours display flux levels for
Lyman-$\alpha$ emission from 0.3 to 8$\times$10$^{-19}$ erg/s/cm$^{2}$/pix$^{2}$, with 
bins of 1.25$\times$10$^{-20}$ erg/s/cm$^{2}$/pix$^{2}$. 
The brightest source seen in the HST image is likely to be a
cluster
galaxy, but no spectroscopic redshift could be determined for it.
} 
\label{System46}
\end{center}
\end{figure}

\item {\bf System 50 (L50a, 588(50b) and L50c):} 
MUSE spectroscopy has allowed us to confirm this spectacular three-image 
system based on the detection of a strong emission-line identified as
Lyman-$\alpha$ for a source at $z =$ 4.274, and seen in the three images available in this
field. Previous photometric redshifts reported by C10 were highly inaccurate (between $z =$ 2.2 
and 3.2). Therefore, these observations allow us to confirm the redshift and nature of this system. 

\end{itemize}

Unfortunately, only a tentative spectroscopic redshift 
could be obtained for the long arcs located southeast in
Fig.\ \ref{finding_chart}, namely 8a and 8b in C10, despite a strong
magnification. These arcs were identified by C10 as two images of the
same source, with photometric and lensing redshifts of 1.8$\pm$0.5 and 2.22,
respectively. Extracted spectra are very noisy, and strong spectral features 
such as [OII]3727 and Lyman-$\alpha$ are not expected within the MUSE
wavelength domain. Here we propose a tentative redshift of $z =$ 2.661, mainly
based on several consistent absorption features, such as CII 1334, 
SiIV 1393 $+$ OIV 1402, FeII 1608, and AlII 1670. A faint emission-line is
also detected in one of the portions of 8a, consistent with CIII]1909, but the
S/N ratio is very low compared to the other line emitters discussed here. A
longer exposure time and further investigation are needed to fully confirm
this redshift.

\subsection{New multiply imaged sources behind A1689}
\label{new_sources}

\begin{figure}
\begin{center}
\includegraphics[width=0.48\textwidth]{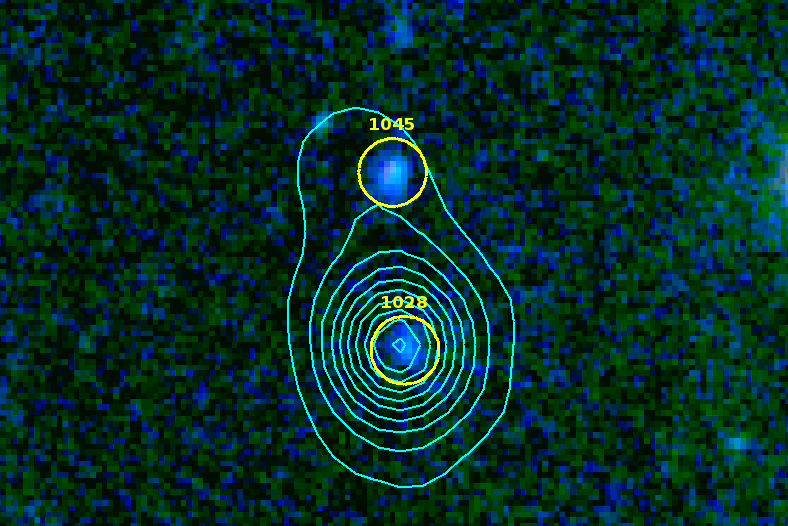}
\caption{Color image of A1689 obtained by combining F475W, F775W, and F850LP
HST/ACS images using a logarithmic scale, 
showing a 6.3\arcsec $\times$ 4.2\arcsec region around 1028
and 1045. 
Circles have diameters of 0.60\arcsec, 
which is close to the FWHM seeing measured in the MUSE data.
Cyan isocontours display flux levels for
Lyman-$\alpha$ emission from 0.5 to 35$\times$10$^{-19}$ erg/s/cm$^{2}$/pix$^{2}$, with 
bins of 2$\times$10$^{-20}$ erg/s/cm$^{2}$/pix$^{2}$. 
} 
\label{System52}
\end{center}
\end{figure}

   As a result of this survey, 14 new background sources were identified
behind A1689, as reported in Table~\ref{catalog}. All of them generate
multiply imaged systems, identified from N$ =$ 51 to 64 (in continuity with
respect to C10). 
In two cases, we were able to spectroscopically confirm
the identification of (at least) two such images: 

\begin{itemize}

\item {\bf System 51 (1339 and LN20):} 
1339 is a highly magnified image of a source at $z =$ 0.775, based on the
identification of several emission-lines, mainly 
[OIII]5007,4959, H$\beta$, H$\gamma$, NeIII, the [OII]3727 doublet, and MgII. 
LN20 is an image extracted within the BCG area and displays the same (main)
emission lines. These two images are consistent with a new three-image system,
with two images located within the MUSE field of view. In this case,
magnifications affecting 1339 and LN20 are $\sim$13 and $\sim$2, respectively,
which agrees well with the relative line fluxes. 
The third image is found as expected in the field that is covered by HST images,
with m(775W)=25.1 and magnification $\sim$5.

\item {\bf System 52 (1028, 1045 and LN12):} 
1028 is a highly magnified image of a source at $z =$ 3.112, based on the
identification of Lyman-$\alpha$ and CIV seen in emission. Lyman-$\alpha$ in
emission is also seen at the same redshift in a region associated with 1045
(in the close neighborhood of 1028), as shown in Fig.\ \ref{System52}.
The system is consistent with a new three-image source, with two images
located within the MUSE field of view, namely 1028 (and its Lyman-$\alpha$
extension 1045), and LN12, which were extracted within the BCG area. 
The third image is identified in the field covered by HST images,
with m(775W)=26.9 and magnification $\sim$4. 

\end{itemize}

Some newly discovered sources in this field deserve a special mention.  

\begin{itemize}

\item {\bf Systems 53, 54, and 55 (LN9, 1404, and 835):}
These three images display a similar redshift z$\sim$3.18. However, based on
the lensing model, they cannot be reconciled with a multiple-image system. LN9
and 1404 are two independent systems with three multiple images each, two of them 
within the MUSE field, namely the image detected and another counter-image within the BCG region $\text{that is about five}$ times fainter for LN9 and de-magnified ($\mu\sim$0.03) for 1404.
A candidate third image could be extracted for system 54 at the expected position in
the field covered by HST images, close to a bright cluster galaxy, with
m(775W)=27.4. In contrast, there is no obvious counterpart for
the faint continuum emission associated with LN9 despite a 1.2
times larger magnification, but the image is expected in the close neighborhood of a
particularly crowded region. 
835 is more likely a double-image system, with a de-magnified counter-image
located within the BCG region.

\item {\bf System 56 (LN10):}
This image displays the largest magnification of the sample ($\mu
\sim$75). It is most likely a new three-image source at z$\sim$3.419, with two images located within the
MUSE field of view. The magnification of the counter-image expected within the
BCG area is $\mu\sim$9, yielding a flux below $\sim$2$\times$10$^{-18}$
erg/s/cm$^{2}$, which is too faint to be detected in this area. 
There is an extremely faint object in the field covered by HST images,
with m(775W)$\sim$28 and magnification $\sim$8, which is a good
candidate to the third image.

\end{itemize}

For all newly discovered galaxies, in addition to those discussed 
above, we checked the consistency of the lensing configuration by identifying
possible counter-image candidates predicted in the field covered by
HST data. For images lacking a detection in the continuum (systems 57,
58, 61, 63, and 64), there is no obvious counter-image in HST data. The only
exception is system 62, for which a faint arclet is detected at the
predicted position, with m(775W)=27.6 and magnification $\sim$15.5, that is, a
larger magnification than in the image seen in the MUSE field. There is no
obvious counter-part for the faint systems 59 and 60. In the first case, the
magnification is expected to be 1.2 times larger, but the image is predicted
in the close neighborhood of a bright cluster galaxy. In the second case, the
counter-image is expected to be $\sim$0.4 magnitudes fainter. 

\subsection{Discussion}
\label{model}

   All background sources detected within the MUSE field of view correspond to 
multiply imaged systems. The spectroscopic images
described above are consistent with this fact. In most cases, they
correspond to three- or five-image systems with at least two 
images located within the MUSE field. The detection of counterparts within the
BCG area, with observed line fluxes expected to be around or below
$\sim$2$\times$10$^{-18}$ erg/s/cm$^{2}$, is simply unfeasible. Therefore, in
the majority of cases presented above, there is no spectroscopic confirmation
of the multiple system, and this is a very common situation in the literature
(i.e., the spectroscopic redshift being available for only one of the
images in a multiple system). However, for sources exhibiting a continuum (i.e., detected
in HST images), we have identified possible counterparts outside the field of
MUSE. 
The mass-model used in this analysis is accurate enough to permit a
good determination of magnification factors and related quantities and
to identify multiple systems. A new revised mass-model,
including all the new multiple images discussed above, will be presented in a
coming paper (Richard et al., in preparation). 


   Compared to the sample of 26 bright well-behaved Lyman-$\alpha$ emitters observed by MUSE
 in the HDFS at $z \ge 2.9$ and displaying extended Lyman-$\alpha$ emission 
\citep[i.e., isolated, non-AGN galaxies][]{2015A&A...575A..75B,  
2015arXiv150905143W},  
our sample is intrinsically fainter, with Lyman-$\alpha$ luminosities ranging between 
40.5$\ltapprox$log(L)[erg.s$^{-1}$]$\ltapprox$42.5
after correction for
magnification. As a result of the relatively short exposure time
and because
these observations were performed close to the extended halo of BCG in the
cluster core, we do not discuss the extended Lyman-$\alpha$
emission around individual sources here. However, such an emission is clearly
observed in at least four cases as described above (systems 1c, 7b, 46a,
and 52;
see Figs.\ \ref{System1} to \ref{System52}).
 
\begin{figure}
\begin{center}
\includegraphics[width=0.50\textwidth]{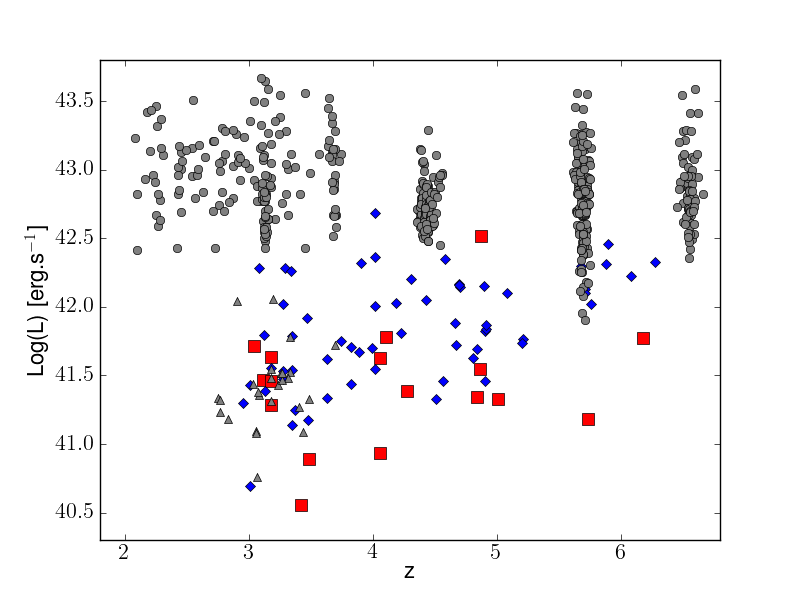}
\caption{Distribution of lens-corrected Lyman-$\alpha$ luminosities for the 17
galaxies found behind A1689 at $3 \le z \le 7$ (red squares), as compared to
\citet[][]{2015A&A...575A..75B}  
(blue diamonds), and 
to a compilation of representative samples of LAE from the literature (gray symbols):
2$\lessapprox$z$\lessapprox$3.8 \citep{2011ApJ...736...31B} (HETDEX Pilot Survey), 
2.7$\lessapprox$z$\lessapprox$3.8 (gray triangles) \citep{2008ApJ...681..856R}, 
z$\sim$4.9 \citep{2009ApJ...696..546S}, 
z$\sim$5.7
\citep{2006PASJ...58..313S,2008ApJS..176..301O,2012ApJ...744..149H,2010ApJ...725..394H}, 
and z$\sim$6.5
\citep{2010ApJ...723..869O,2010ApJ...725..394H,2011ApJ...734..119K}. 
} 
\label{Luminosity_comparison}
\end{center}
\end{figure}

   When we compare the distribution of lens-corrected Lyman-$\alpha$
luminosities for the 17 galaxies found behind A1689 at
3$\ltapprox$z$\ltapprox$7 to a compilation of representative 
spectroscopic samples from the literature (see Fig.\ \ref{Luminosity_comparison}),
the complementarity between classical blank-field and lensing studies is
obvious. Lyman-$\alpha$ emitters
observed by MUSE in lensing clusters nicely extend the accessible domain in
luminosity between one and two orders of magnitude beyond classical LAE
surveys. The luminosities reached here are comparable with those attained by 
\citet[][]{2008ApJ...681..856R} 
in their $\sim$90h exposure with VLT/FORS2
(gray triangles in Fig.\ \ref{Luminosity_comparison}). 
The present sample also reaches fainter limits than the survey
conducted on the HDFS by \citet[][]{2015A&A...575A..75B},  
in particular at z$\gtrapprox$4, which is due to the specific pointing covering the
highest magnification regime in this particularly efficient cluster. 

   A high magnification value also translates into a small effective
surface or volume covered by the survey on the corresponding source planes. To
evaluate this effect, we computed the effective volume covered by the
survey in the different redshift bins from z$\sim$3.5 to z$\sim$6.5, with
redshift slices of $\Delta z =$ 0.5. A mask was used to remove the areas
contaminated by the central BCGs, where faint sources typically below a few 
$\sim\times$10$^{-18}$ erg/s/cm$^{2}$ could not be detected. This mask
accounts for $\sim$11\% of the total surface. 
We integrated the effective volume outside the mask using the same approach as in 
\citet[][]{2006A&A...456..861R},  
that is, the surface of each pixel was reduced by the corresponding magnification
(depending on position and redshift). 
We used Lenstool and the same lensing model as in Sect.\ \ref{cluster} to
derive the magnification maps for each redshift slice. 
The total effective covolume obtained in this way in the field of view of 
MUSE is $\sim$145 Mpc$^{3}$ at 3$\ltapprox$z$\ltapprox$7, whereas it is
$\sim$2300 Mpc$^{3}$ in a blank field. 

\begin{figure}
\begin{center}
\includegraphics[width=0.50\textwidth]{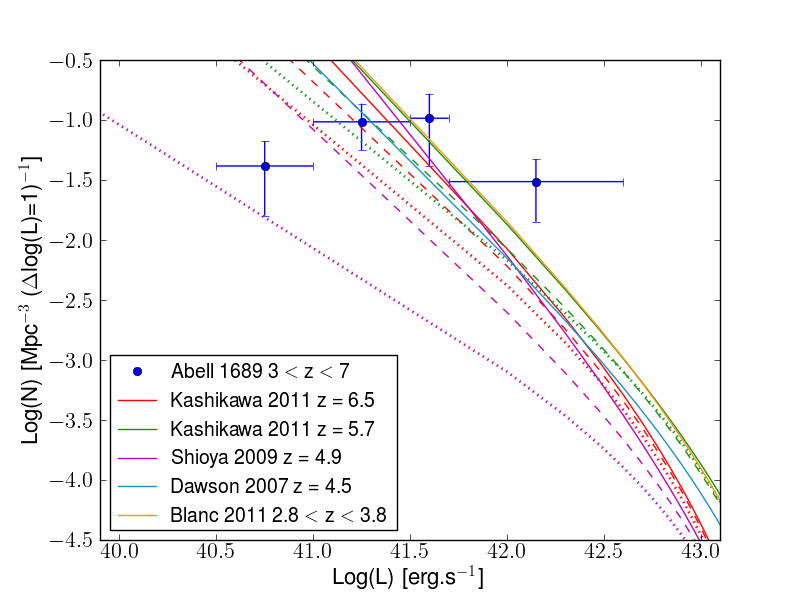}
\caption{Comparison between the density of sources observed at 
$3 \le z \le 7$ behind A1689 and expectations based on the extrapolation
of the LF toward the low-luminosity regime covered by the present survey. References are
given in the figure for different redshifts. Solid lines display the steepest
slopes adopted for the LF fit: $\alpha = -1.7$ for 
\citet[][]{2011ApJ...734..119K} and  
\citet[][]{2011ApJ...736...31B}, 
$\alpha = -2.0$ for 
\citet[][]{2009ApJ...696..546S}, and 
$\alpha = -1.6$ for 
\citet[][]{2007ApJ...671.1227D}. 
Dashed and dotted lines correspond to 
$\alpha = -1.5$ and -1.3 for \citet[][]{2011ApJ...734..119K} and 
$\alpha = -1.5$ and -1.0 for \citet[][]{2009ApJ...696..546S}. 
Error bars include Poisson noise statistics and field-to-field variance 
(see text). 
}
\label{LF_high-z_Lya}
\end{center}
\end{figure}

   Figure\ \ref{LF_high-z_Lya} displays the comparison between the density of
sources obtained in this way within the whole 3$\ltapprox$z$\ltapprox$7
interval, and expectations based on the extrapolation of the luminosity
function (LF) toward the low-luminosity regime covered by the present survey. 
Luminosity bins were defined to keep at least three sources per bin,
and they are independent. Given the small number of sources, we did not
introduce any correction for incompleteness at this 
stage, knowing that incompleteness is expected to affect in particular the 
faintest bins in Fig.\ \ref{LF_high-z_Lya}. 
Error bars in this figure include Poisson noise statistics that
affect the
number of galaxies found in a given luminosity bin, as well as an estimate of
the field-to-field variance, added in quadrature. Field-to-field variance was
estimated using the public cosmic variance calculator by 
\citet[][]{2008ApJ...676..767T}.  
The error budget is clearly dominated by Poisson noise in this case. 

   Given the intrinsic low luminosities in our sample, we are particularly
sensitive to the value of the slope parameter $\alpha$, for which a constant
value is assumed in other surveys presented in
Figs.\ \ref{Luminosity_comparison} and\ \ref{LF_high-z_Lya}, where the fit
is only sensitive to L$^{*}$ and the normalization. 
Despite these caveats discussed above regarding the small size of this sample, the density of 
intrinsically faint sources observed in this field is roughly consistent with
the steepest values used by the different authors to fit their data, 
namely $\alpha$ $\le$ -1.5, and inconsistent with flatter values of $\alpha$. 
Any correction for completeness in the faintest bins will exacerbate this trend. 
This result still needs further investigation to be
confirmed, using larger samples of LAE in this low-luminosity domain. 
A full analysis of the LF for the faint Lyman-$\alpha$ emitters behind
lensing clusters will be presented elsewhere in a forthcoming paper. 

\section{Conclusions}
\label{Conclusions}

  To summarize, these are the main results we obtained:

\begin{itemize}
    
\item Based on guided-and-manual and automated (blind) approaches, a census of
sources was obtained in the field of MUSE, most of which are cluster
galaxies. The spectroscopic census is complete out to m(775W) $<$ 21.5. 
The spectroscopic sample includes 63 cluster galaxies with secure redshifts
(49 of them with m(775W) $<$ 21.5). 
The velocity distribution of cluster galaxies is more complex than the
smooth Gaussian distribution found for this cluster on larger scales, with a
main cluster component with $\sigma =$ 1200 km/s, and the signature of the NE
group component, with $\sigma =$ 600 km/s. This demonstrates that 
MUSE data are able to catch the small-scale
substructure in particularly dense environments, a difficult exercise for usual
multi-slit spectrographs. This result supports the mass-model method adopted by L07
and used throughout the paper to retrieve lens-corrected quantities.

\item We are able to confirm 12 images
for background sources that correspond to 7 different lensed galaxies between $z =$ 0.95 and 5.0 of the 28 different lensed galaxies known or suspected within the MUSE
field of view. In addition, 14 new galaxies were spectroscopically identified 
in this area thanks to MUSE data, with redshifts ranging between 0.8 and 6.2, 
the majority of them identified based on their Lyman-$\alpha$ emission. 
All background sources detected within the MUSE field of view correspond to 
multiply imaged systems, in most cases three- or five-image systems with at least two 
images located within the MUSE field.

\item In total, 17 sources are found at $z \ge 3$ based on their Lyman-$\alpha$
emission, with Lyman-$\alpha$ luminosities ranging between 
40.5$\ltapprox$log(Ly$\alpha$)$\ltapprox$42.5 after correction for
magnification. Four of these systems display extended Lyman-$\alpha$ emission.
Our sample is intrinsically fainter than the usual samples available in the 
literature, in particular at $z \ge 4$, and comparable to the faintest samples
currently available at lower redshifts 
\citep[e.g., the sample Lyman-$\alpha$ emitters observed 
by MUSE in the HDFS at $z \ge 2.9$ ][]{2015A&A...575A..75B,  
2015arXiv150905143W}.  
Given the intrinsically low luminosities, this sample is particularly
sensitive to the slope of the LF toward the faintest-end.
The density of sources obtained in this survey is roughly consistent with
a steep value of $\alpha \le -1.5$, although this result still needs
further investigation. 

\end{itemize}

   These results illustrate the efficiency of MUSE in characterizing
lensing clusters on one hand, and studying faint and distant populations
of galaxies on the other hand. In particular, our current survey of lensing
clusters is expected to provide a unique census of sources responsible for the
reionization in a limited representative volume at z$\sim$4-7.

\begin{acknowledgements}
We thank O. Czoske for providing unpublished spectroscopic data on A1689.
Part of this work was supported by the French CNRS. 
This work has been carried out thanks to the support of the OCEVU Labex
(ANR-11-LABX-0060) and the A*MIDEX project (ANR-11-IDEX-0001-02) funded by the
"Investissements d'Avenir" French government program managed by the ANR. 
Partially funded by the
ERC starting grant CALENDS (JR, VP, BC),  
and the Agence Nationale de la recherche bearing the
references ANR-09-BLAN-0234 and ANR-13-BS05-0010-02 (FOGHAR). 
Based on observations made with ESO Telescopes at the La Silla
Paranal Observatory under programme ID 60.A-9100(B). We thank
all the staff at Paranal Observatory for their valuable support during the
commissioning of MUSE.  
Also based on observations made with the NASA/ESA Hubble Space Telescope 
(PID 9289). 
\end{acknowledgements}


\bibliographystyle{aa}  
\bibliography{A1689_lens_paper} 



\clearpage
\onecolumn
\landscape
\centering
\setlength\topmargin{1cm}
\begin{table}[p]
\caption{\label{catalog} Catalog of images identified behind the
  lensing cluster A1689 within the field of view of MUSE. 
All images corresponding to the same source are grouped 
together and separated by horizontal lines.}
\begin{tabular}{lcccccccrl}
\hline
  ID &  ID & RA (J2000) & DEC (J2000) & z$_{spec}$ & Line & Flux & m(775W) & $\mu$ & Comments  \\
     & (1) &            &             &   (2)     &  (3) & 10$^{-18}$ erg/s/cm$^{2}$ & AB &  & 
Other lines detected/ extraction issues  \\
\hline
1339&  51a  & 13:11:28.620 & -1:20:15.52 & 0.7752 & [OII]5007 & 177.2$\pm$1.3 &24.33 & 12.9 
  & [OIII]5007,4959, Hb, Hg, He, Hd, NeIII, [OII], MgII \\
LN20&  51b   & 13:11:29.334 & -1:20:25.08 & 0.7752 & [OII]5007 &  27.2$\pm$1.1 & 25.65& 2.3
  & [OIII]5007,4959, Hb \\
\hline
L26b & 26b & 13:11:31.361 & -1:20:23.76 & 0.9594 & [OII]3727 & 5.0$\pm$1.0 & 25.62& 10.1
 & [OII]3727,3729 \\
\hline
750 & 4a & 13:11:32.212 & -1:20:55.91 & 1.1649 & [OII]3727 & 23.4$\pm$0.2  &24.16 &  16.8
&  \\
518 & 4b & 13:11:30.546 & -1:21:10.64 & 1.1649 & [OII]3727 & 11.5$\pm$0.4 &23.03 &  11.3
&  \\
L4e & 4e & 13:11:29.877 & -1:20:27.93 & 1.1649 & [OII]3727 & 4.5$\pm$0.2 & - &  1.0
& Faint line extracted within the BCG area \\
\hline
L17a& 17a& 13:11:30.697 & -1:20:23.43 & 2.6614 & CIII]1909 & 10.2$\pm$0.2 & 24.33 & 4.7
&    \\
\hline
619 & 1c & 13:11:29.809 &  -1:21:06.01 & 3.0445 & Lyman-$\alpha$ & 40.1$\pm$0.4 &24.62 &  6.2
&  \\
LN23 & 1c(ext) & 13:11:29.858 & -1:21:06.59 & 3.0445 & Lyman-$\alpha$ & 43.4$\pm$0.4  & 27.89 &  6.3
&  emission blob associated to 619/1c \\
LN24 & 1c(ext) & 13:11:29.756 & -1:21:06.02 & 3.0445 & Lyman-$\alpha$ & 38.3$\pm$0.4  & 27.82 &  6.3
&  emission blob associated to 619/1c  \\
L1f & 1f & 13:11:29.892 & -1:20:37.06 & 3.0445 & Lyman-$\alpha$ & 17.2$\pm$0.4 & 24.91 & 1.9
&  Extracted within BCG spectrum \\
\hline
1028 & 52a & 13:11:31.582 & -1:20:39.31 & 3.1124 & Lyman-$\alpha$ & 72.2$\pm$0.4 & 26.86 & 20.8
&  Lines detected: Lyman-$\alpha$ and CIV1550 \\
1045&  52a(ext)  &  13:11:31.591 & -1:20:37.79 & 3.1124 & Lyman-$\alpha$ & 23.7$\pm$0.4 & 26.24 & 20.5
&  extended emission associated to 1028 \\
LN12 &  52b   &  13:11:30.472 & -1:20:35.07 & 3.1124 & Lyman-$\alpha$ & 22.8$\pm$0.4 & - &  3.0
& Emission seen on 2D spectra \\
\hline
LN9  &  53  &  13:11:30.199 &  -1:20:57.83 & 3.1803 & Lyman-$\alpha$ & 35.2$\pm$0.4 & 27.82 & 7.3
&   \\
\hline
1404 &  54  & 13:11:31.123 &  -1:20:12.08 & 3.1813 & Lyman-$\alpha$ & 9.4$\pm$0.5 & 26.83 & 4.4 
&  \\
\hline
835 &  55 & 13:11:30.720 &  -1:20:53.12 & 3.1813 & Lyman-$\alpha$ & 18.2$\pm$0.4 & 26.22 & 5.6  
&  \\
\hline
LN10&  56  &  13:11:28.887 &  -1:20:51.63& 3.4188 & Lyman-$\alpha$ & 25.7$\pm$0.4 & 26.77 & 75.0
&   \\ 
\hline
944 & 46a & 13:11:31.703 & -1:20:45.80 & 3.4837 & Lyman-$\alpha$ & 5.0$\pm$0.4 &  24.91& 22.4
&  \\ 
LN22 & 46a(ext) & 13:11:31.686  &-1:20:46.93 & 3.4837 & Lyman-$\alpha$ & 13.7$\pm$0.4 & - &  19.6 
& emission-line blob associated to 944/46a \\
\hline
LN26 & 57 & 13:11:28.899  &-1:21:08.88 & 4.0555 & Lyman-$\alpha$ & 11.9$\pm$0.4 & - &  21.9
&  \\ 
\hline
LN13 & 58  & 13:11:29.067 & -1:20:57.49 & 4.0555 & Lyman-$\alpha$ & 18.4$\pm$0.4 & - &  6.8
&  \\ 
\hline
LN14 & 59  & 13:11:31.004  &-1:20:14.28 & 4.1048 & Lyman-$\alpha$ & 16.2$\pm$0.4 & 26.99 & 4.4 
&  \\ 
\hline
L50a & 50a &  13:11:32.610 & -1:20:42.08 & 4.2746 & Lyman-$\alpha$ & 34.1$\pm$0.9 & 26.67 & 25.2
&  \\ 
588 & 50b & 13:11:31.055 & -1:21:07.65 & 4.2746 & Lyman-$\alpha$ & 23.2$\pm$0.5 &26.27&  10.0
&   \\ 
L50c& 50c & 13:11:31.672 & -1:20:12.26 & 4.2746 & Lyman-$\alpha$ & 12.9$\pm$0.5 & 27.88 & 5.1
&  \\ 
\hline
 LN25 & 60 & 13:11:28.656 & -1:20:54.62 &4.8444 & Lyman-$\alpha$ &  8.8$\pm$0.5 & 26.970 & 9.6
&  \\ 
\hline
LN15  & 61  & 13:11:30.347 & -1:21:07.82 & 4.8680 & Lyman-$\alpha$ & 8.0$\pm$0.4 & - & 5.6
&  \\ 
\hline
1379& 7b & 13:11:30.709 &  -1:20:12.56  &4.8742 & Lyman-$\alpha$ & 59.0$\pm$0.5 & 24.2 & 4.4
&  \\
1401&  7b(ext) &13:11:30.723 &  -1:20:12.27 & 4.8742  & Lyman-$\alpha$ & 58.2$\pm$0.5 & 26.19 & 4.4   
 &  emission-line associated to 1379/7b \\ 
LN16  & 7c &   13:11:30.107 & -1:20:25.94 & 4.8742  & Lyman-$\alpha$ & 28.6$\pm$0.4 & - & 0.7
 &  counter-image extracted within BCG area \\
\hline
LN17&  62 & 13:11:28.995 & -1:21:07.30 & 5.0120 & Lyman-$\alpha$ & 4.9$\pm$0.4 & - & 5.9
&  \\ 
\hline
LN18&  63 &  13:11:31.413 & -1:20:45.34 & 5.7382 & Lyman-$\alpha$ & 6.1$\pm$0.4 & - & 14.5
&  \\ 
\hline
LN19& 64 & 13:11:30.257 & -1:21:00.78 & 6.1763 & Lyman-$\alpha$ & 8.4$\pm$0.5 & - & 6.0
&  \\ 
\hline
\end{tabular}
\tablefoot{\\
(1) Multiple-image identification with N $\le$ 50 come from \citep{2010ApJ...723.1678C}. 
The same convention is used for the new multiple images identified by MUSE. Multiplicity or
substructure within the same image is indicated by the extension ``ext''. 
\\ 
(2) Redshifts based on ref. vacuum wavelengths \\
(3) Main emission line 
}
\end{table}
\endlandscape
\twocolumn


\end{document}